\documentclass[preprint,prb, amsmath,amssymb,aps,superscriptaddress,nobibnotes]{revtex4-1}
\usepackage{CJK}
\usepackage{graphicx}
\usepackage{color}

\usepackage{natbib}
\setcitestyle{comma}
\usepackage{dcolumn} 
\usepackage{bm}     
\usepackage{amsfonts}
\usepackage[bookmarks=false,colorlinks,citecolor=red]{hyperref}
\usepackage{amsmath}

\usepackage[version=4]{mhchem}
\usepackage{siunitx}
\usepackage[normalem]{ulem}

\newcommand{\bfl}{\begin{flushleft}}
\newcommand{\efl}{\end{flushleft}}

\newif\ifshowcomments\showcommentstrue

\usepackage{xr-hyper}
\makeatletter
\newcommand*{\addFileDependency}[1]{% argument=file name and extension
  \typeout{(#1)}
  \@addtofilelist{#1}
  \IfFileExists{#1}{}{\typeout{No file #1.}}
}
\makeatother

\newcommand*{\myexternaldocument}[1]{%
    \externaldocument{#1}%
    \addFileDependency{#1.tex}%
    \addFileDependency{#1.aux}%
}
\myexternaldocument{Supplement}

\newcommand{\LTO}{LiTi$_2$O$_4$}

\begin{document}
\include{MyCommand}

\title{
Unconventional polaronic ground state in superconducting \LTO{}
}
\author{Zubia Hasan}
\thanks{These authors contributed equally to this work.}
\affiliation{Department of Physics, Harvard University, Cambridge, MA, USA}
\author{Grace A. Pan}
\thanks{These authors contributed equally to this work.}
\affiliation{Department of Physics, Harvard University, Cambridge, MA, USA}
\author{Harrison LaBollita}
\affiliation{Department of Physics, Arizona State University, Tempe, AZ, USA}
\author{Austin Kaczmarek}
\affiliation{Laboratory of Atomic and Solid-State Physics, Cornell University, Ithaca, NY, USA}
\author{Suk Hyun Sung}
\affiliation{The Rowland Institute, Harvard University, Cambridge, MA, USA}
\author{Shekhar Sharma}
\affiliation{Department of Physics, Arizona State University, Tempe, AZ, USA}
\author{Purnima P. Balakrishnan}
\affiliation{NIST Center for Neutron Research, National Institute for Standards and Technology, Gaithersburg, MD, USA}
\author{Edward Mercer}
\affiliation{Department of Physics, Northeastern University, Boston, MA, 02115, USA}
\affiliation{Quantum Materials and Sensing Institute, Northeastern University, Burlington, MA, 01803 USA}
\author{Vivek Bhartiya}
\affiliation{National Synchrotron Light Source II, Brookhaven National Laboratory, Upton, NY 11973, USA}
\author{Zaher Salman}
\affiliation{PSI center for Neutron and Muon Sciences, 5232 Villigen PSI, Switzerland}
\author{Thomas Prokscha}
\affiliation{PSI center for Neutron and Muon Sciences, 5232 Villigen PSI, Switzerland}
\author{Andreas Suter}
\affiliation{PSI center for Neutron and Muon Sciences, 5232 Villigen PSI, Switzerland}
\author{Alexander J. Grutter}
\affiliation{NIST Center for Neutron Research, National Institute for Standards and Technology, Gaithersburg, MD, USA}
\author{Mirian Garcia-Fernandez}
\affiliation{Diamond Light Source, Harwell Campus, Didcot OX11 0DE, UK.}
\author{Ke-Jin Zhou}
\affiliation{Diamond Light Source, Harwell Campus, Didcot OX11 0DE, UK.}
\author{Jonathan Pelliciari}
\affiliation{National Synchrotron Light Source II, Brookhaven National Laboratory, Upton, NY 11973, USA}
\author{Valentina Bisogni}
\affiliation{National Synchrotron Light Source II, Brookhaven National Laboratory, Upton, NY 11973, USA}
\author{Ismail El Baggari}
\affiliation{The Rowland Institute, Harvard University, Cambridge, MA, USA}
\author{Darrell G. Schlom}
\affiliation{Platform for the Accelerated Realization, Analysis, and Discovery of Interface Materials (PARADIM), Cornell University, Ithaca, NY, USA}
\affiliation{Department of Materials Science and Engineering, Cornell University, Ithaca, New York 14853, USA}
\author{Matthew R. Barone}
\affiliation{Platform for the Accelerated Realization, Analysis, and Discovery of Interface Materials (PARADIM), Cornell University, Ithaca, NY, USA}
\author{Charles M. Brooks}
\affiliation{Department of Physics, Harvard University, Cambridge, MA, USA}
\author{Katja C. Nowack}
\affiliation{Laboratory of Atomic and Solid-State Physics, Cornell University, Ithaca, NY, USA}
\author{Antia S. Botana}
\affiliation{Department of Physics, Arizona State University, Tempe, AZ, USA}
\author{Brendan D. Faeth}
\affiliation{Platform for the Accelerated Realization, Analysis, and Discovery of Interface Materials (PARADIM), Cornell University, Ithaca, NY, USA}
\author{Alberto de la Torre }
\thanks{Correspondence should be addressed to: \href{mailto:a.delatorreduran@northeastern.edu}{a.delatorreduran@northeastern.edu}; \href{mailto:mundy@fas.harvard.edu}{mundy@fas.harvard.edu}}
\affiliation{Department of Physics, Northeastern University, Boston, MA, 02115, USA}
\affiliation{Quantum Materials and Sensing Institute, Northeastern University, Burlington, MA, 01803 USA}
\author{Julia A. Mundy}
\thanks{Correspondence should be addressed to: \href{mailto:a.delatorreduran@northeastern.edu}{a.delatorreduran@northeastern.edu}; \href{mailto:mundy@fas.harvard.edu}{mundy@fas.harvard.edu}}
\affiliation{Department of Physics, Harvard University, Cambridge, MA, USA}
\date{\today}
\maketitle

\noindent \textbf{Geometrically frustrated lattices can display a range of correlated phenomena, ranging from spin frustration and charge order to dispersionless flat bands due to quantum interference.  One particularly compelling family of such materials is the half-valence spinel Li$B_2$O$_4$ materials. On the $B$-site frustrated pyrochlore sublattice, the interplay of correlated metallic behavior and charge frustration leads to a superconducting state in LiTi$_2$O$_4$ and heavy fermion behavior in LiV$_2$O$_4$.  To date, however, LiTi$_2$O$_4$ has primarily been understood as a conventional BCS superconductor despite a lattice structure that could host more exotic groundstates.  Here, we present a multimodal investigation of LiTi$_2$O$_4$, combining ARPES, RIXS, proximate magnetic probes, and ab-initio many-body theoretical calculations. Our data reveals a novel mobile polaronic ground state with spectroscopic signatures that underlie co-dominant electron-phonon coupling and electron-electron correlations also found in the lightly doped cuprates. The cooperation between the two interaction scales distinguishes \LTO{} from other superconducting titanates, suggesting an unconventional origin to superconductivity in \LTO{}. Our work deepens our understanding of the rare interplay of electron-electron correlations and electron-phonon coupling in unconventional superconducting systems. In particular, our work identifies the geometrically frustrated, mixed-valence spinel family as an under-explored platform for discovering unconventional, correlated ground states.}

The interplay between electron-electron correlations and electron-phonon coupling has been of long-standing interest in understanding superconductivity. In traditional Bardeen-Cooper-Schrieffer (BCS) superconductors, Coulomb repulsion between electrons is thought to screen electron-phonon coupling and reduce the superconducting transition temperature $T_c$\cite{matthias_chapter_1957}. In unconventional superconductors, however, there can be a more complex interplay between electron-phonon coupling and electron-electron interactions. Strong correlations can further strengthen the electron-phonon coupling and enhance $T_c$ \cite{capone_strongly_2002,yuan_correlation-enhanced_2022,nomura2015unified}. In certain cases, as explored for cuprates \cite{scalapino2012common}, the energy scales of strong electron-electron correlations can overwhelm those of electron-phonon interactions, leading to charge and spin fluctuations that compete with or enhance superconductivity\cite{van_loon_competing_2018}. Thus, understanding how electronic correlations and electron-phonon coupling interact in superconductors is important in identifying new families of unconventional superconductors and ascertaining the mechanisms behind high temperature superconductivity.   

\begin{figure*}
    \centering
    \includegraphics[width = \columnwidth]{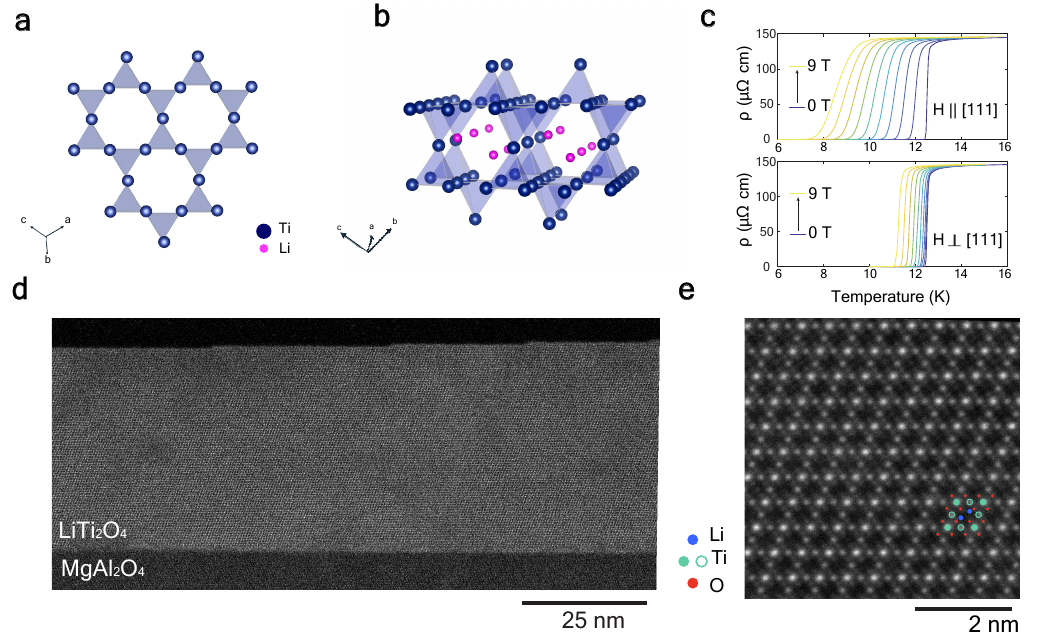}
    \caption{\textbf{Structural and electrical characterization of \LTO{}} \textbf{a}, The (111) plane of \LTO{} showing the titanium kagome sublattice inherent to the spinel structure. \textbf{b}, Side-view of the (111) plane shown in \textbf{a}, highlighting the three-dimensional nature of the geometric frustration in \LTO{}. \textbf{c},  Resistivity $\rho$ vs temperature of \LTO{} with magnetic field (0 to 9 T) parallel and perpendicular to the sample. \textbf{d}, zoomed out HAADF-STEM image of \LTO{} showing uniform crystallinity over a large area. \textbf{d}, Zoomed-in HAADF-STEM image of \LTO{} overlaid with the corresponding atoms.}
    \label{fig:structure}
\end{figure*}

Although superconductivity was discovered in \LTO{} before the cuprates\cite{johnston_superconducting_nodate}, it remains a unique instance of a spinel superconductor and with the highest $T_c$ for a mixed-valence titanate. In this crystal structure, $d^{0.5}$ titanium atoms form a pyrochlore sublattice, comprised of alternating planes of Kagome and triangles (Fig. \ref{fig:structure}a and \ref{fig:structure}b). This geometric construction can give rise to flat bands due to quantum interference, which has led to strong electron correlations in other systems\cite{wakefield_three-dimensional_2023}. 
Still, superconductivity in \LTO{} has primarily been considered as phonon-mediated BCS-like supported by specific heat, tunneling spectroscopy, and muon spin rotation measurements~\cite{sun_magnetic_2004,jin_anomalous_2015,wu_magnetic_1994}. In contrast, anomalous magnetotransport behavior and scanning tunneling spectroscopy experiments have suggested the presence of spin fluctuations, orbital ordering, and pseudo-gap likes states in \LTO{}, all of which are typically associated with unconventional superconductors and strongly correlated materials~\cite{jin_anomalous_2015,okada_scanning_2017,xue_fourfold_2022}. Interestingly, previous theoretical studies of \LTO{} have suggested non-BCS-like superconducting mechanisms, such as bipolaronic~\cite{edwards_study_1984,alexandrov_bipolaronic_1981} or resonating valence bond (RVB) superconductivity~\cite{anderson_resonating_1987,anderson_resonating--valence-bond_1987}. While there have been speculations about the role of electron-electron correlations \cite{oda_electron-phonon_1994,sun_magnetic_2004}, a more detailed investigation into the strength and nature of various correlations is needed to explore the notion of  "unconventional superconductivity" in LiTi$_2$O$_4$. 

Here, we reveal the complex interplay between electron-phonon coupling and electron-electron interactions in \LTO{}. We use molecular-beam epitaxy (MBE) to synthesize epitaxial thin films of superconducting \LTO{}, enabling a detailed spectroscopic investigation using resonant inelastic x-ray scattering (RIXS) and angle-resolved photoemission spectroscopy (ARPES). The combination of the element-specific sensitivity to structural and local excitations of RIXS together with the unique capability of ARPES to reveal energy and momentum dependence of the quasiparticle self-energy enable us to provide a comprehensive description of the low energy physics of LiTi$_2$O$_4$. We observe evidence of strong electronic correlations and signatures of strong electron-phonon coupling. The cooperation between these two interaction scales gives rise to a novel polaronic ground state - also found in weakly doped cuprates\cite{shen_missing_2004} - that dynamically localizes titanium states in LiTi$_2$O$_4$. This interpretation is further supported by our theoretical calculations, which can reproduce some of the spectral features in our ARPES data when considering electron correlations or electron-phonon interactions separately but are unable to reproduce polaronic phenomena associated with the interplay of both interactions. Our work thus challenges the notion of phonon-dominated BCS superconductivity in LiTi$_2$O$_4$ by revealing the complex correlations present in this material, hearkening comparisons to cuprate-like physics. 
\\
\bfl {\bf \large Results} \efl
\noindent\textbf{Synthesis and characterization of superconducting \LTO{}}
\\
Thin films of LiTi$_2$O$_4$ were grown via reactive oxide MBE on (111)-oriented MgAl$_2$O$_4$ substrates (see Methods). As shown in Fig. \ref{fig:structure}c, the films display a superconducting transition with $T_c \sim$ 12.5 K and a residual resistivity ratio (RRR) of $\sim5.3$ (Fig. S3)  consistent with previous reports of the highest quality materials\cite{johnston_superconducting_nodate, jin_anomalous_2015}. Figure \ref{fig:structure}d,e show high-angle annular dark field scanning transmission electron microscopy (HAADF-STEM) images of LiTi$_2$O$_4$, attesting to the high structural quality of our films (also see Supp. Fig. S1).

\begin{figure*}
    \centering
    \includegraphics[width = \columnwidth]{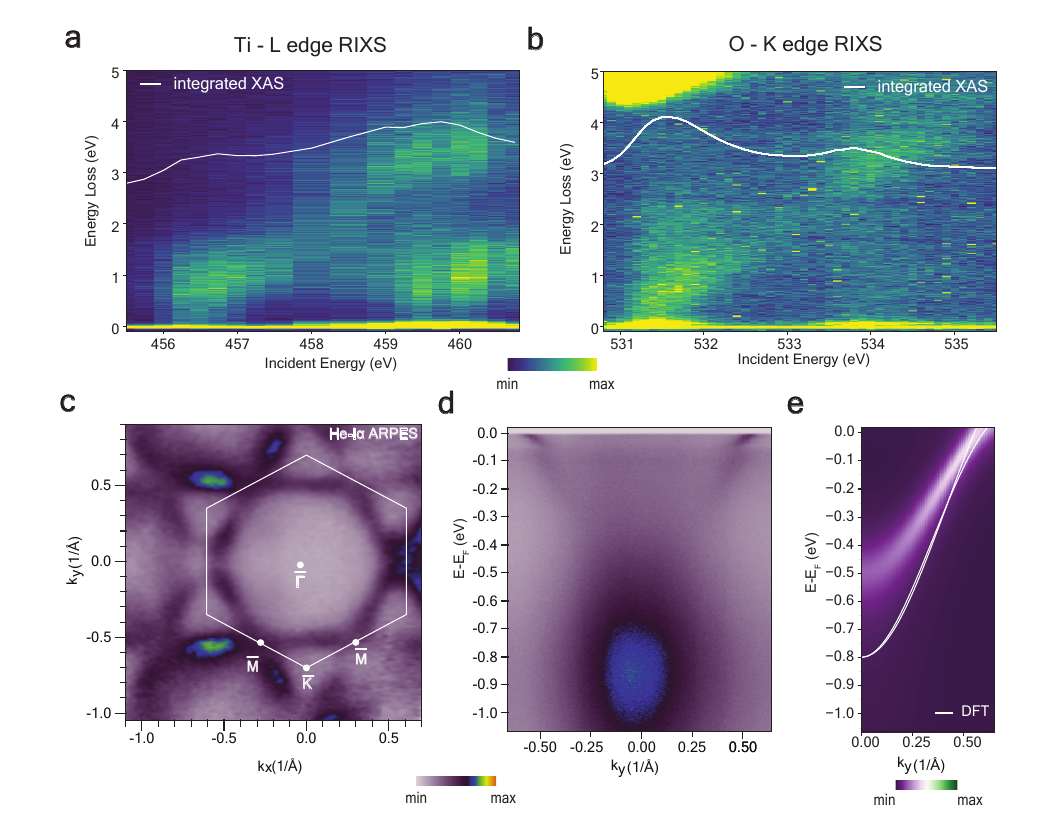}
    \caption{\textbf{Electron-electron correlations in \LTO{}} \textbf{a}, RIXS intensity at the Ti-$L_{3}$ edge. \textbf{b}, RIXS intensity at the O-$K$ edge. \textbf{c}, Iso-energy map at the Fermi level taken with He-I$\alpha$ showing a hexagonal Fermi surface. The white hexagon indicates the first Brillouin zone. \textbf{d}, Experimental band structure in the $\bar{K}-\bar{\Gamma}-\bar{K}$ direction. \textbf{e}, DMFT and DFT (white) comparison along $\bar{\Gamma}-\bar{K}$ indicating the mass renormalization due to electron-electron correlations.}
    \label{fig:electron_electron}
\end{figure*}

Additionally, we confirm the onset of superconductivity and probe the superconducting order parameter using scanning superconducting quantum interference device (SQUID) microscopy, which locally measures magnetic susceptibility (Fig. S4). The measured magnetic susceptibility can be used to extract the temperature dependence of the reduced London penetration depth (Supp. Mat. Sec. S3), which can be well-fit by a fully gapped order parameter, consistent with $s$-wave-like pairing. The observation of the 2-D thin film limit of superconductivity, the high RRR, as well as a homogeneous diamagnetic response from scanning SQUID demonstrates the high structural quality of our films, ensuring that further measurements probe the intrinsic behavior of \LTO{}. 
\\
\\
\noindent\textbf{Electron-electron correlations in \LTO{}}
\\
Poised with high-quality thin films, we first probe the electronic structure using elemental-specific RIXS measurements.  Figure \ref{fig:electron_electron}a shows the RIXS intensity of \LTO{} near the Ti-$L_{3}$ edge. Our x-ray absorption spectroscopy (XAS) data (Fig. S13), displays resonant features associated with the mixed valence Ti$^{4+}$ and Ti$^{3+}$ states. Given the nearly featureless RIXS spectrum expected from a $d^0$ state \cite{geondzhian_large_2020}, we conclude that the RIXS map in Fig. \ref{fig:electron_electron}a, reflects a $d^1$ titanium occupation state, which given the lack of charge order in \LTO{} can only be dynamically populated \cite{vibronic_RIXS_KIC}. Our RIXS data is characterized by broad Raman-like features centered at $E_{loss} = 1.5$ eV and $E_{loss} = 4$ eV. These excitations, along with the suppressed fluorescence contribution, are unexpected, given that the itinerant charge carriers in our metallic and superconducting samples are proposed to have a dominant Ti-3$d$ band character near the Fermi level \cite{massidda_electronic_1988,satpathy_electronic_1987}. We note that our RIXS data resembles that of MgTi$_2$O$_4$  \cite{li_evolution_2023} despite \LTO{} lacking large static local trigonal distortions of the TiO$_6$ octahedra and Ti-Ti dimerization (Supp. Fig. S1 and Supp. Fig. S17). The qualitative similarities between the RIXS energy maps in metallic \LTO{} and Mott-insulating systems like MgTi$_2$O$_4$ are suggestive of the presence of localized excitations in \LTO{}.

In addition to the Raman-like excitations, we observe spectral weight associated with a fluorescence contribution with a linear dispersion in $E_i$ that extends below $E_{loss} \leq 1.5$~eV. This resembles the superposition of localized and delocalized excitations observed in negative charge transfer insulators \cite{bisogni2016ground}. Moreover, O-$K$-edge RIXS, shown in Fig. \ref{fig:electron_electron}b, also shows Raman-like and fluorescence contributions below the charge transfer gap, $\Delta \approx 4.5 $~eV. This is the same energy as the intra-$dd$ excitations at the Ti-$L_3$ edge, suggesting strong hybridization between titanium and oxygen carriers. While the hybridization between Ti-$3d$ and O-$2p$ states has been previously theorized \cite{massidda_electronic_1988}, the RIXS data is the first instance of direct observation of strong titanium-oxygen hybridization. Moreover, the presence of fluorescence contributions at the O-$K$-edge near the elastic line (zero energy loss) suggests a larger contribution of oxygen carriers to electronic transport in \LTO{} than that expected from previous density of states (DOS) calculations \cite{massidda_electronic_1988}. Thus, our RIXS data suggest the presence of charge localization, Ti-$3d$ and O-$2p$ hybridization, and electron-electron correlations in \LTO{}.

Electron-electron correlations are also observed in \textit{in-situ} ARPES measurements on \LTO{}. Figure \ref{fig:electron_electron}c shows the Fermi surface of \LTO{} from our ARPES measurements. Here, $k_x$ and $k_y$ are chosen to lie along the [11$\bar{2}$] and [$\bar{1}$10] directions respectively. The Fermi surface is characterized by large electron pockets with hexagonal symmetry centered at $\Gamma$, pushing toward the zone boundary in proximity to a Lifshitz transition, which has led to enhanced electronic interactions near the Fermi surface for other superconductors \cite{shi2017enhanced}. A more detailed discussion of the Fermi surface and its $k_z$ dependence is given in Supp. Fig. S5, S6 and in the Supp. Mater. Sec S4. 

To analyze the electronic interactions, we look at high-symmetry cuts of the band structure of \LTO{}. In Fig. \ref{fig:electron_electron}d,e, we compare our experimental band structure along $\bar{\Gamma}-\bar{K}$ to the momentum-resolved spectral function obtained from density-functional theory plus dynamical mean-field theory (DFT+DMFT). The non-interacting band dispersion is shown in white in Fig.~\ref{fig:electron_electron}e for reference. The DFT+DMFT calculation reproduces the experimental dispersion near $\Gamma$, capturing the large effective mass renormalization compared to the bare DFT band -- a clear indication of strong electron correlations in \LTO{}. The calculated cyclotron mass (Supp. Mater. Fig. S9) yields $\frac{m_{eff}}{m_{bare}} = $ 2.53 $\pm$ 0.4. This method of calculating the renormalization agrees with previous reports of the effective mass of \LTO{} by other optical probes\cite{ohsawa_origin_2020,zhao_pseudo-dielectric_2023}. Additionally, we observe a broad Gaussian-like, intense incoherent spectral weight centered at $E_B \approx 0.9$~eV that resembles the lower Hubbard band of lightly doped cuprates \cite{myasnikova2018relaxation}. Intriguingly, the spectral weight disappears in the data taken by Ne-I$\alpha$ photon energy (16.85 eV), but the ``water-fall" like features remain (Supp. Fig. S6). We note this feature cannot be reproduced by DMFT calculations but is consistent with a previous photoemission spectroscopy measurement of \LTO{}, where it was interpreted as a signature of polaronic behavior \cite{edwards_study_1984}. Nonetheless, our combined ARPES and RIXS data provide the first observation of significant electronic correlations in LiTi$_2$O$_4$.
\\
\\

\begin{figure}
    \centering
    \includegraphics[width = 1\columnwidth]{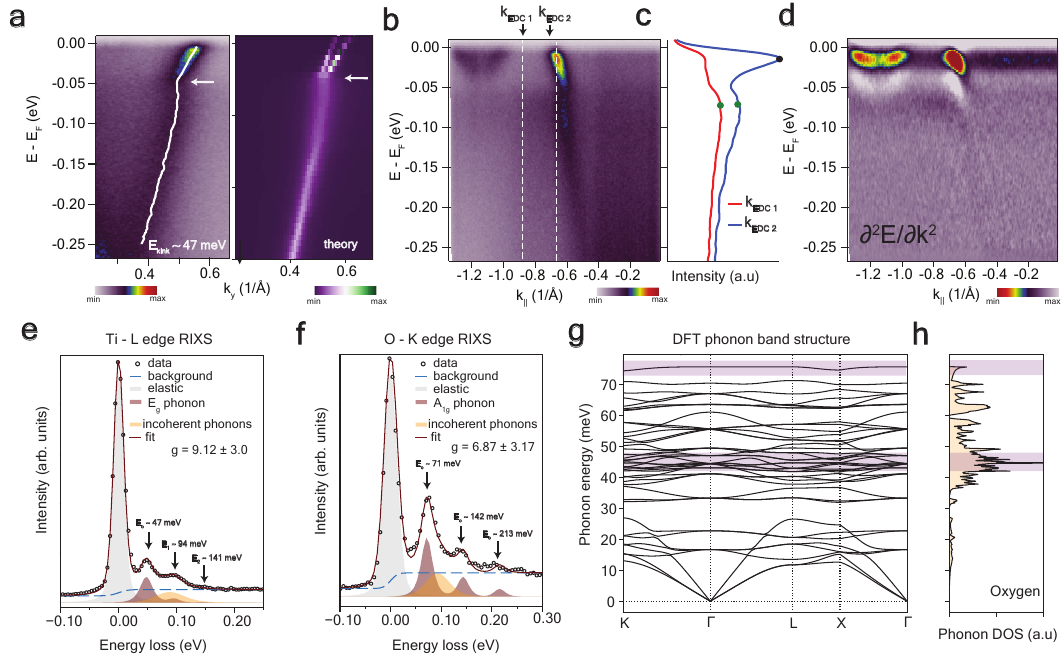}
    \caption{\textbf{Electron-phonon coupling in \LTO{}} \textbf{(a)} Comparison of the $\bar{\Gamma}-\bar{K}$ ARPES data to density-functional perturbation theory (DFPT) highlighting the kink feature at 47 meV at $\bar{\Gamma}-\bar{K}$ at $ T = 7$~ K. \textbf{b}, Energy - momentum ARPES spectra along the $\bar{\Gamma}-\bar{K}-\bar{M}$ direction. \textbf{c}, EDC cuts at two different $k_{||}$ highlighted in \textbf{b} with dotted white lines. Black and green dots correspond to the top of the two EDC poles.\textbf{d}, Second-derivative plot from \textbf{b}, highlighting a dispersionless spectral intensity at the second pole denoted by a black arrow in \textbf{c} also observable in the raw data in \textbf{b}. A spectral gap is also prominent at the same energy scale where the dispersionless feature interrupts the main band. \textbf{e}, Quasi-elastic RIXS excitations shown at $E_{inc} = 460$~ eV for the Ti-$L_{3}$ edge. \textbf{f}, Quasi-elastic RIXS excitations shown at $E_{inc} =  531.5$~eV for the O-$K$ edge. The total intensity is a fit (red solid line) to a Voigt peak (grey shedding), a background step function (blue dashed lines), and Ament Model fit to the phonon spectra (maroon) in \textbf{e} and \textbf{f}. \textbf{g} and \textbf{h}, Calculated phonon band structure and total oxygen phonon density of states respectively. The purple boxs highlights the energy scale of the $A_{1g}$ and $E_g$ modes, respectively, in both plots.}    
    \label{fig:electron_phonon}
\end{figure}

\noindent\textbf{Electron-phonon coupling in \LTO{}}
\\
While electronic correlations are important for unconventional superconductivity, we also observe signatures of electron-phonon coupling in the electronic structure of \LTO{}. Consistent with previous ARPES reports \cite{fujisawa2023imaging}, we observe a `kink' at $E_{B} \approx$ 46 meV as shown in Fig. \ref{fig:electron_phonon}a.  This feature is ascribed to an $E_g$ oxygen phonon mode by tunneling spectroscopy measurements \cite{gilmore_description_2021} and inelastic neutron scattering \cite{green_lattice_1997}, confirmed by calculations of the spectral function including the self-energy from only electron-phonon interactions (Supp. Mat. Sec. S5). We extract a band renormalization, $\lambda_{tot} = 1.80(1)$ three times higher than $\lambda_{e-ph} = 0.65$ determined by specific heat measurements \cite{sun_magnetic_2004} and ab-initio calculations \cite{oda_electron-phonon_1994}. The extracted band renormalization value from our ARPES data is comparable to that from electronic specific heat measurements, where the larger value is attributed to enhancement from electron-electron correlations or spin fluctuations  \cite{heintz_superconductivity_1989,satpathy_electronic_1987,mccallum_superconducting_1976}. For a more detailed discussion on the calculation of $\lambda_{tot}$ as well as its temperature and momentum dependence, see Sec. Mater. Sec S4 and Fig. S7. 

Figure \ref{fig:electron_phonon}b shows an energy-momentum cut along the $\bar{\Gamma}-\bar{K}-\bar{M}$ direction (cut overlaid on the Brillouin zone is shown in Fig. S8a). Here we observe incoherent spectral weight at $E_{B} = 46 \pm 10$~meV at all $k_{||}$. To examine this non-dispersive feature, we present two energy dispersive cuts (EDC) in Fig. \ref{fig:electron_phonon}c, $k_{EDC 1}$ which cuts across momentum value devoid of bands and $k_{EDC 2}$ which cuts across the $k_{F}$ of the main \LTO{} band. The momentum position of these two cuts are indicated by two black arrows pointing to dotted white lines in Fig. \ref{fig:electron_phonon}b. We can see the distinct signature of this flat feature in the EDC of $k_{EDC 1}$ as an abrupt and sharp increase in spectral intensity at $\Omega$ $=$ 46 $\pm 10$ meV (green dot) also observed in $k_{EDC2}$ at the same energy scale. These observations contrast the expected EDC for a non-interacting picture where, in the absence of bands, we would not expect any additional poles in the EDC, unlike what is seen at $k_{EDC1}$. The second derivative plot (Fig. \ref{fig:electron_phonon}c) highlights this feature and additionally shows a ``spectral gap" where the flat feature interrupts the main band electron pocket. We note that the flat feature, in addition to the spectral gap, can also be seen clearly in our raw data in Fig. \ref{fig:electron_electron}d. Spectral signatures such as these are consistent with the formation of intra-unit cell small polarons \cite{hohenadler_spectral_2003} that have been observed in other correlated materials \cite{kang_holstein_2018,Fulleride_2023}. 

Our RIXS measurements also corroborate the presence of strong electron-phonon coupling and multiphonon processes in \LTO{}. Figure \ref{fig:electron_phonon}e shows the RIXS spectra at the Ti-$L_3$ and O-$K$ edge, respectively. As highlighted by black arrows in Fig. \ref{fig:electron_phonon}e and f, both RIXS spectra are characterized by prominent quasi-elastic peaks that are approximately equally spaced and monotonically decay with $E_{loss}$, thus resembling the harmonic progression of the multiphononic processes. The low-energy excitations shown in the titanium spectra are well fit by the Ament model \cite{ament_determining_2011}, which considers a single non-dispersive phonon mode coupled to the electronic structure with strength $g$ (See Supp. Mater. Sec S6 for more details on the Ament model fit). At the Ti-$L_{3}$ edge, the Ament model in conjunction with a broad Gaussian peak centered at  $E_{loss} = 90$~meV provides a good fit for a mode centered at $\Omega = 47 \pm 2$ meV, in agreement with the oxygen $E_g$ mode observed in ARPES. We interpret this broad Gaussian peak to represent multiple incoherent phononic excitations that cannot be resolved within the energy resolution of the instrument, also observed in  MgTi$_2$O$_4$\cite{li_evolution_2023}. 

The coupling strength, $g_{Ti}= 9.1 \pm 3.0$, far exceeds those of other titanates \cite{moser_electron-phonon_2015,fatale_hybridization_2016}, and is on par with some cuprates measured via RIXS \cite{braicovich_determining_2020} further supporting strong electron-phonon coupling in LiTi$_2$O$_4$. Moreover, the observation of a pure oxygen $E_g$ mode at the titanium edge is indicative of strong hybridization between the O-$2p$ and the Ti-$3d$ orbitals. Similar to the Ti-$L_3$ edge, the three excitations at the O-$K$ edge can also be well fit by the Ament model in conjunction with a broad Gaussian peak. The fit gives us the first three harmonics of a phonon mode centered at $\Omega = 71 \pm 1$ meV with $g_{O} = 6.89 \pm 3.17$. We assign this mode to the oxygen $A_{1g}$ phonon, which is known to be strongly coupled to the $E_g$ mode \cite{oda_electron-phonon_1994}. We note that our DFT calculations of the phonon band structure (Fig. \ref{fig:electron_phonon}g) show the modes around $\Omega \approx 46 $ meV and $\approx 75$ meV to be relatively nondispersive, supporting our choice of Ament model fitting for the quasi-elastic peaks at the Ti-$L_{3}$ and O-$K$ edge. 

\bfl {\bf \large Discussion} \efl

\noindent Our combined spectroscopic data indicates that the effects of electron-electron correlations and electron-phonon coupling cannot be disentangled in \LTO{}. Moreover, \textit{ab-initio} calculations (Fig. \ref{fig:electron_electron}e and Fig. \ref{fig:electron_phonon}a) treating electron-phonon coupling and electron-electron correlations separately fail to capture the most salient features of our data, namely the existence of localized excitations in \LTO{} (Fig. \ref{fig:electron_electron}a) and the presence of a second pole in the photoemission self-energy (Fig. \ref{fig:electron_phonon}c). While the combined theoretical description of these two energy scales remains challenging and a matter of active research \cite{PhysRevMaterials.7.093801}, there is growing evidence that the excitation spectrum, band structure and transport properties of quantum materials such as SrVO$_3$ \cite{abramovitch2024respective} and superconductors like alkalli-fullerides \cite{Fulleride_2023} and the cuprates\cite{he2018rapid} can only by understood through the interplay of both interaction scales. We note that while competing interactions is generally associated with unconventional superconductivity, our scanning SQUID data is consistent with a conventional s-wave order parameter (Supp. Sec. S3). This naturally also raises the question as to how the phonon - mediated attractive interaction - essential for $s$-wave superconductivity - is robust against or potentially enhanced by the repulsive electronic correlations apparent in \LTO{}. Within this context, \LTO{} departs from SrTiO$_3$ \cite{wang_tailoring_2016} and other superconducting titanates \cite{zhang_enhanced_2017,yoshimatsu_superconductivity_2017} where electron-phonon coupling is understood to be the dominant interaction.

\begin{figure}
    \centering
    \includegraphics[width = 1\columnwidth]{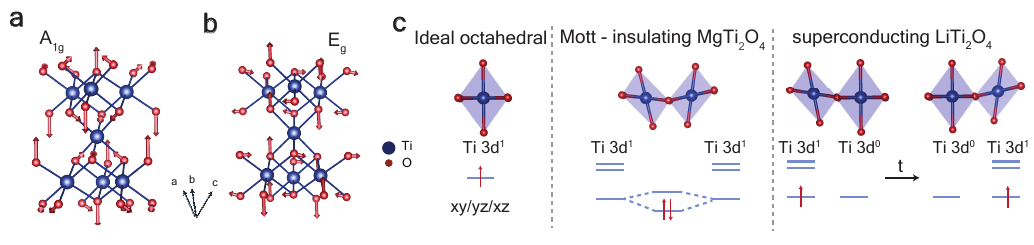}
    \caption{\textbf{Polaron formation in \LTO{}}. \textbf{a,b}, DFT calculated A$_{1g}$ and E$_{g}$ oxygen phonon modes in \LTO{}. \textbf{c}, An ideal TiO$_{6}$ octahedra (left) with a highly degenerate $d^{1}$ state in a cubic crystal field. In MgTi$_2$O$_4$, a local trigonal distortion driven by E$_{g}$ and A$_{1g}$ phonons is required to achieve a dimerized ground state (middle). Dynamic symmetry-reducing local lattice fluctuations associated with electron hopping leading to polaron formation in \LTO{} (right)}.     
    \label{fig:polaron_formation}
\end{figure}

We finally comment on the possible mechanism that enables the strong interplay of electron-electron and electron-phonon interactions in \LTO{}. In this context, it is instructive to consider the $d^{0.5}$ occupation state in \LTO{} in the context of the $d^1$ member of the AB$_2$O$_4$ family, MgTi$_2$O$_4$. In MgTi$_2$O$_4$, a trigonal distortion driven by $E_g$ and $A_{1g}$ oxygen phonons \cite{popovic2003phonon} lifts the orbital degeneracy of the Ti $3d^1$ state in a perfect cubic octahedral environment (Fig. \ref{fig:polaron_formation} a, b). The combination of this local distortion with the formation of Ti-Ti dimers at $T=260$~K\cite{popovic2003phonon} leads to a tetragonal, dimerized unit cell and the formation of a correlated insulating state (Fig. \ref{fig:polaron_formation}b) in MgTi$_2$O$_4$. In the $d^{0.5}$ mixed-valence state the energy gain from dimerization is reduced in favor of a charge-ordered ground state, such as in mixed-valence spinel CuIr$_2$S$_4$\cite{radaelli2002formation}. However, no static symmetry breaking due to orbital or lattice ordering has been reported in \LTO{}\cite{jin_anomalous_2015}, possibly due to the geometric frustration inherent in the spinel structure. This raises the question of how the $d^1$ state, with its associated local distortion, is accommodated in \LTO{}. We speculate that in \LTO{} with an occupation $d^{0.5}$ in the absence of charge order, two titanium ions share an electron dynamically, such that their occupation state fluctuates between $d^1$ and $d^0$. In this scenario, the transiently occupied $d^1$ site has an associated local symmetry reduction due to dynamic lattice fluctuations of the $E_g$ and $A_{1g}$ oxygen phonons, as seen in MgTi$_2$O$_4$ \cite{yang2020two}. As the electron moves to the unoccupied $d^0$ site, it drags the local distortion with it, leading to the formation of a polaronic state in which charge motion and lattice distortions are coupled (Fig. \ref{fig:polaron_formation}b). 

Our data reflects the formation of such a small mobile polaronic state in \LTO{}. First, our ARPES data shows a strong coupling ($\lambda > 1$) to a non-dispersive $E_g$ phonon mode. The ratio of the dominant $E_g$ phonon mode energy ($\Omega = 46$~meV), to the bare electron bandwidth ($t \approx 800$~meV), places \LTO in the adiabatic limit ($\Omega/t << 1$) for which the condition $\lambda>1$ is expected for a small polaron ground state \cite{capone2000small}. Second, the observation of the same mode ($E_{g}$ oxygen mode) at the Ti-$L_{3}$ edge indicates a tightly hybridized state between the localized $d^{1}$ titanium electrons and the deformed lattice. Third, we observe large incoherent spectral weight at $E = 800$~meV, which cannot be accounted by either of our density-functional perturbation theory (DFPT) or DMFT calculations. We note that polaronic behavior has been previously discussed for \LTO{} based on the anomalous transport behavior of off-stoichiometric samples \cite{watanabe1984semiconducting} and previous photoemission and reflectance spectroscopy data \cite{edwards_study_1984,harrison_study_1984}. Notably, the metallic behavior of \LTO{} at all temperatures contrasts with the localized small Holstein polaron behavior of other titanates \cite{yang_intrinsic_2013,lakkis_metal-insulator_1976} where a low-temperature insulating state occurs due to reduced lattice mobility. However, in the light regime ($m^* < 10m$), small polarons have been theoretically predicted to be mobile \cite{davenport2012mobile,hague2007superlight}. Polaron delocalization involves a crawling-like motion in which an electron is transiently delocalized over two neighboring sites \cite{holstein1959studies}, similar to the mechanism suggested here for the $d^{0.5}$ state in \LTO{} (Fig. \ref{fig:polaron_formation}b) in which the dynamical lattice fluctuations are coupled to electron hopping between $d^1$ and $d^0$ states. We note that this mobile polaron formation requires an intricate balance between electron-electron correlations and electron-phonon coupling. Dominant electron-electron correlations may choose a charge-ordered ground state \cite{radaelli2002formation} while dominant electron-phonon coupling can lead to bi-polaron formation as suggested for other titanates, like Ti$_4$O$_7$\cite{lakkis_metal-insulator_1976}. Thus, our data suggests a serendipitous balancing of electron-electron correlations with electron-phonon coupling in this material.

Although \LTO{} has long been thought of as a well-studied BCS superconductor consistent with phonon-dominated superconductivity as in other titanates\cite{zhang_enhanced_2017,yoshimatsu_superconductivity_2017} - our work forces us to reconsider the notion of conventional superconductivity in \LTO{}. Here, a re-examination of \LTO{} using  ARPES and RIXS measurements uncovers strong hybridization between titanium and oxygen, considerable electron-electron correlations, and co-existing electron-phonon coupling. In mixed valence \LTO{}, this balance of energy scales results in a novel mobile polaronic ground state indicative of a unique balance of charge delocalization, electron-electron correlations, and electron-phonon coupling. \LTO{} becomes a model system to explore predictions of enhancement of superconductivity due to the cooperative effect of electron-electron correlations and electron-phonon coupling in the quarter-filled Hubbard Hamiltonian \cite{PhysRevResearch.2.023006}. Moreover, the proximity of superconductivity to an orbitally ordered phase, the correlated behavior, and the strong hybridization between titanium and oxygen highlights a rich competition between energy scales found only in special classes of quantum materials\cite{}. Our work expands our understanding of superconductivity in $d^{0.5}$ systems and demonstrates how mixed-valence spinel-oxide structures can host correlated physics, thus broadening our search of material families that can host such correlated phenomena.

\newpage

\bfl {\bf Acknowledgements} \efl
\noindent We thank useful discussions with G. Grissonnanche, M. R. Norman, Y. Wang, F. Baumberger, and J. Sous. This research is primarily supported by the National Science Foundation, Division of Materials Research, under Award No. DMR-2339913.  Materials growth and photoemission studies were supported by the Platform for the Accelerated Realization, Analysis, and Discovery of Interface Materials (PARADIM) under NSF Cooperative Agreement No. DMR-2039380.  All nanofabrication work was performed at Harvard University's Center for Nanoscale Systems (CNS), a member of the National Nanotechnology Coordinated Infrastructure Network (NNCI), supported by the National Science Foundation under NSF Grant No. 2025158.  Z.H. and G.A.P. acknowledge support from the Paul \& Daisy Soros Fellowship for New Americans.  G.A.P. acknowledges additional support from the NSF Graduate Research Fellowship Grant No. DGE-1745303. A.K. and K.C.N. acknowledge support from the Air Force Research Laboratory, Project Grant FA95502110429.  S.H.S. and I.E.B. acknowledge support from the Rowland Institute at Harvard University.  J.A.M. acknowledges support from the Packard Foundation and the Sloan Foundation. This research used beamline 2-ID of the National Synchrotron Light Source II which is a US DOE Office of Science Facility operated for the DOE Office of Science by Brookhaven National Laboratory under contract no. DE-SC0012704. HL and ASB acknowledge support from NSF Grant No. DMR-2323971. The $\mu$SR experiments were performed at the Swiss Muon Source, S$\mu$S, Paul Scherrer Institute, Villigen, Switzerland.  

\bfl {\bf Author Contributions} \efl
\noindent Z.H., G.A.P., C.M.B. and J.A.M. synthesized the thin films with assistance from M.R.B and D.G.S. Electrical transport measurements were performed and analyzed by G.A.P. Scanning transmission electron microscopy was performed by S.H.S. and I.E. ARPES measurements were performed by Z.H., G.A.P. and B.D.F. ARPES analysis was done by Z.H. and B.D.F. with support by A.d.l.T. The RIXS measurements were performed by Z.H., E.M., S.H.S., I.B. and A.d.l.T. with support from V.B., V.B. and J.P. and M.G.F. and K.Z. RIXS analysis was done by Z.H., S.H.S and E.M under the supervision of A.d.l.T. $\mu$SR measurements were performed by A.J.G and P.P.B with support from A.S., Z.S., T.P. Analysis of $\mu$SR data was done by A.J.G and P.P.B. Scanning SQUID measurements were performed by A.K. and K.C.N. H.L, S.S. and A.S.B performed the DFT, DMFT, and electron-phonon calculations. J.A.M. and A.d.l.T. conceived and guided the study. Z.H., G.A.P., J.A.M., and A.d.l.T. wrote the manuscript with contributions and discussion from all authors.

\bfl {\bf Competing Interests} \efl
\noindent The authors declare no competing interests.
\clearpage
\newpage

\clearpage
\newpage

\newpage
\bfl {\bf \large Methods} \efl
\bfl {\bf Synthesis of \LTO{} thin films via MBE} \efl
We used reactive oxide molecular beam epitaxy (Veeco GEN 10) to synthesize thin film LiTi$_2$O$_4$ on untreated MgAl$_2$O$_4$ (111) substrates (CrysTec GmbH).  The lithium and titanium fluxes were matched and set to $\sim$1.5 $\times$ 10$^{13}$ atoms/cm$^2 \cdot$ s as measured by a quartz crystal microbalance (QCM).  We obtained the lithium flux by flowing O$_2$ during the QCM process and forming Li$_2$O, as the low atomic mass of elemental lithium yields small frequency changes on the QCM below the detection sensitivity except for at very high fluxes.  For the deposition process, we heated the substrates with a 10.6 $\mu$m CO$_2$ laser to 825 $^{\circ}$C in a pressure of 1.0 - 2.0 $\times$ 10$^{-7}$ molecular O$_2$ over a chamber background pressure of $\sim$5.0 $\times$ 10$^{-8}$.  We co-deposited the lithium and titanium sources for one hour. After deposition, we shut off the O$_2$ flow to prevent oxidation of titanium to the 4+ state and cooled at a rate of 100$^{\circ}$C/min.

\bfl {\bf Structural characterization} \efl
Annular dark field (ADF) STEM was performed on Thermofisher Scientific (TFS) Themis (operated at 200 keV, convergence semi-angle 18.9 mrad, ADF collection angle: 36--200 mrad). A small ADF inner collection angle was used to increase light element sensitivity. Electron transparent TEM samples were prepared using TFS Helios DualBeam FIB/SEM. 

\bfl {\bf Transport measurements} \efl
We performed all electrical transport measurements in a Hall bar geometry using evaporated Cr/Au (7 nm/100 nm) contacts with the Hall channel defined by a diamond scribe along the [11-2] crystallographic direction of the MgAl$_2$O$_4$ (111) substrate.  We loaded the samples in 9 T Dynacool Physical Property Measurement System (PPMS) and used AC lock-in techniques at $\sim$15 Hz.

\bfl {\bf Resonant inelastic X-ray scattering } \efl
The XAS and RIXS spectra were collected at the SIX beamline at NSLS-II and the I21 beamline of Diamond Light Source. The Ti-$L_3$ edge was measured at SIX and I21, while the O-$K$ edge spectra was acquired at I21. The Ti-$L_{3}$ energy map is collected at 2$\theta$ = 150$^{\circ{}}$, while the O-$K$ energy map is collected at 2$\theta$ = 90$^{\circ{}}$. All RIXS spectra are collected at 2$\theta$ = 90$^{\circ{}}$ at grazing incidence of $\theta$ = 20$^{\circ{}}$ unless indicated otherwise. The energy resolution was fixed between 20 to 25 meV at both SIX and Diamond, respectively. All data was measured at a base temperature of 22 K at both beamlines. 

\bfl {\bf Angle resolved photoemission spectroscopy } \efl
All \textit{in situ} photoemission measurements were conducted by immediately transferring the samples through a UHV manifold (P $<$ 2 $\times$ 10$^{-9}$ Torr) to a measurement chamber immediately following film growth. ARPES measurements were taken with a Scienta Omicron DA30-L electron analyzer equipped with a Fermion Instruments BL1200s multi-gas discharge lamp using He-I photons at 21.2 eV, Ne-I photons at 16.85 eV, and He-II photons at 40.2 eV. The base pressure in the ARPES system is maintained during measurements at pressures lower than \smash{5 $\times$ 10$^{-11}$}~Torr. ARPES measurements were performed at a temperature of 7 K and a nominal experimental energy resolution of 10 meV unless otherwise indicated.

\bfl {\bf Muon spin relaxation measurements} \efl
Low-energy muon spin-relaxation (LE-$\mu$SR) measurements were done using the LEM instrument at the Swiss muon source. An applied field of 10 mT was applied transverse to the $\mu^+$s spin polarization direction. For more details on the $\mu$SR measurements, refer to Supp. Mat. Sec. S7.

\bfl {\bf Data Availability} \efl
\noindent The data supporting the findings of this study are available from the corresponding authors upon reasonable request. 

\bibliography{bib.bib}

\end{document}